# Fabrication of Ultra-Thick Masks for X-ray Phase Contrast Imaging at Higher Energy

*Alessandro Rossi  Ian Buchanan  Alberto Astolfo  Martyna Michalska  Daniel Briglin  Anton Charman Daniel Josell  Sandro Olivo\*  Ioannis Papakonstantinou\**

Alessandro Rossi
Photonic Innovations Lab, Department of Electronic & Electrical Engineering, University College London, Torrington Place, London, WC1E 7JE UK

Dr. Ian Buchanan
Department of Medical Physics and Biomedical Engineering, University College London, Gower Street, London WC1E 6BT, United Kingdom

Dr. Alberto Astolfo
Department of Medical Physics and Biomedical Engineering, University College London, Gower Street, London WC1E 6BT, United Kingdom

Dr. Martyna Michalska
Manufacturing Futures Lab, Department of Mechanical Engineering, University College London, Queen Elizabeth Olympic Park, London E20 3BS, U.K.

Dr. Daniel Briglin
Nikon X-Tek Systems, Tring Business Centre, Tring, Hertfordshire, U.K.

Anton Charman
Nikon X-Tek Systems, Tring Business Centre, Tring, Hertfordshire, U.K.

Dr. Daniel Josell
Materials Science and Engineering Division, National Institute of Standards and Technology, Gaithersburg, MD 20899

Prof. Sandro Olivo
Department of Medical Physics and Biomedical Engineering, University College London, Gower Street, London WC1E 6BT, United Kingdom
a.olivo@ucl.ac.uk

Prof. Ioannis Papakonstantinou
Photonic Innovations Lab, Department of Electronic & Electrical Engineering, University College London, Torrington Place, London, WC1E 7JE UK
i.papakonstantinou@ucl.ac.uk

Keywords:  *X-ray phase contrast imaging, superconformal Au deposition, stepped parameters etching, phase sensitivity, non-destructive testing*

X-ray phase contrast imaging (XPCI) provides higher sensitivity to contrast between low absorbing objects that can be invisible to conventional attenuation-based X-ray imaging. XPCI's main application has been so far focused on medical areas at relatively low energies (< 100 keV). The translation to higher energy for industrial applications, where energies above 150 keV are often needed, is hindered by the lack of masks/gratings with sufficiently thick gold septa. Fabricating such structures with apertures of tens of micrometers becomes difficult at depths greater than a few hundreds of micrometers due to aspect ratio dependent effects such as anisotropic etching, and preferential gold (Au) deposition at the top of the apertures. In this work, these difficulties are overcome by Deep Reactive Ion Etching optimized by a stepped parameters approach and bismuth-mediated superconformal filling of Au, ultimately resulting in 500 µm deep silicon masks filled with Au at bulk density. The obtained masks, tested in an Edge Illumination XPCI system with a conventional source and a photon-counting detector, show good agreement with simulations at different energy thresholds. They also demonstrate a higher phase sensitivity for highly absorbing objects when compared to lower aspect ratio masks, proving their potential for industrial non-destructive testing.

## 1  Introduction

In the last three decades, X-rays phase contrast imaging has proven it can boost image contrast compared to conventional X-ray imaging using the phase shift induced by objects instead of their attenuation[1–5] During this time, researchers optimized several imaging approaches to detect phase effects, which can be categorized into two overarching groups: interferometric (e.g., Talbot-Lau Interferometry)[6, 7] and refraction-based (e.g., Edge Illumination) methods.[8, 9] Most approaches rely on a combination of absorption and phase gratings to modulate the X-ray beam to generate resolvable signals on the detector plane, ultimately determining the image quality of the system. To date, XPCI has shown great promise for several applications, primarily in biomedical imaging while targeting relatively small and low-Z specimens, with some notable exceptions such as the trial on lung imaging underway in Munich.[10–14] An effective translation to higher energies (> 100 keV) would enable a more widespread application of XPCI to non-destructive testing (NDT) in industrial applications, where tube voltages of 160 kVp or higher are commonly used. This could benefit cultural heritage investigations, security scans, as well as some medical areas requiring higher X-ray energies (e.g., CT).[15–19] However, high energy XPCI has been so far hindered by the lack of high-quality absorption gratings (sometimes referred to as 'masks') with depths sufficient to attenuate harder X-rays. For instance, while only ≈ 10 μm of Au are needed to absorb 90% of a monochromatic beam at 10 keV, 640 μm are required to absorb the same beam fraction at 150 keV; fabricating gratings with the latter thickness and a pitch of a few tens of micrometers, i.e., roughly with aspect ratio (AR) > 25, is a challenging task.

To date, a combination of X-ray or optical lithography and electrodeposition is often used to fabricate X-ray gratings and optical devices. However, X-ray lithography is costly and requires access to synchrotron facilities, while small periods are hard to achieve with optical lithography, and the attainable AR is usually limited in both cases.[20] To overcome these limitations, etching is required to fabricate gratings with higher ARs and/or sub-micrometer periods. Currently the dominant etching methods are Deep Reactive Ion Etching (DRIE) and Metal Assisted Chemical Etching (MACE), a dry and a wet method, respectively.[21, 22] The former's main issues (tapering, bowing, lagging) increase with etching depth, due to the sluggish diffusion of the reactants and their by-products in higher AR features, leading to non-vertical etching.[23, 24] To circumvent these AR-dependent effects, DRIE can be performed at cryogenic temperatures, which however requires specialized and more expensive equipment, or utilizing thorough parameter optimization.[25, 26] Conversely, MACE relies on catalysis of the substrate's oxidation and dissolution by a noble metal previously patterned on it, which sinks in the etched trenches during the process.[27] This leads to a highly anisotropic process depending on the metal stability and its adherence to the substrate, which can be disrupted by e.g., a torque applied by an unbalanced rate across the metal pattern or by the coalescence of gaseous products into bubbles. These mechanisms can tilt, bend, fracture or even delaminate the metal pattern thereby tilting, spreading, and eventually inhibiting the etching process.[28] Since MACE is carried out in solution, capillary forces also act on the etched lamellae upon drying, bending them and causing lamellae agglomerates.[29, 30] To circumvent this, vapour-phase MACE has been recently introduced, usually using customized equipment, and post-etching drying methods can be used to counter drying capillary forces.[30, 31]

The long penetration depths of hard X-rays necessitate use of a dense material in the trenches to achieve significant absorption or refraction. Typically, Au is used due to its high atomic number (Z) and its capacity to be electroplated. However, due to the low conductivity and wettability of typical substrate materials (e.g., silicon and photoresists), deposition of a conductive seed layer is needed first, usually by atomic layer deposition (ALD) or a physical vapor deposition (PVD) process such as magnetron sputtering or thermal evaporation.[30, 32, 33] Electroplating high AR structures can be challenging due to gradients of the metal ion concentration caused by diffusional transport limitations. Strategies to accommodate these limits include the use of very low current densities, pulsed currents, or, as used in this study, additive mediated plating resulting in bottom-up, void-free, superconformal filling.[30, 32, 34, 35] Alternatively, in the MACE framework, the metal catalyst at the trench bottom can also induce a bottom-up Au plating without preliminary seed layer deposition.[36]

These fabrication challenges constrain the attainable AR, and thus depth, of X-ray gratings, so far mostly limited to 300 μm with AR ≈ 30, thereby limiting phase-based imaging to lower energies (maximum of 80 keV).[37] Importantly, the ratio between the unit decrement of the real and imaginary parts ($\delta/\beta$) of the refractive index increases with the photon energy. This indicates an even higher sensitivity of XPCI vs conventional attenuation that is yet barely investigated.[38]

Only recently, examples have appeared where Edge Illumination (EI) XPCI was used at > 100 kVp for security scans and for NDT, with proof-of-concept of multi-modal imaging at 150 kVp.[18, 19, 39] EI-XPCI's robustness and versatility results from its inherent achromaticity and reduced dependency on source size and beam coherence. It is based on the use of a pair of masks modulating the beam into an array of individual, physically separated beamlets transmitted by the masks' X-ray-transparent lamellae. These characteristics make EI a promising technology for high-energy XPCI.[9]

This work demonstrates a route for fabricating silicon (Si) masks filled with Au for EI with periods of 75 µm and 98 µm, apertures of 21 µm and 28 µm, and depths of ≈ 500 µm. Periods and apertures are maintained as in the thinner masks currently in use at Nikon X-Tek Systems, thereby allowing a direct comparison. Both MACE and DRIE were examined for the fabrication. However, the former was unsuccessful due to the large width of the etched trenches. A highly anisotropic etching was successfully achieved with DRIE by dynamically adapting the parameters according to the current AR – similarly to an approach referred to in the literature as "ramped-parameter" DRIE.[25] It is here applied, using discrete steps, to the fabrication of hard X-ray masks for the first time. Finally, the samples were sputter-coated with a thin bilayer of chromium (Cr) and Au to allow for Au bottom-up filling by the Bi-catalyzed electrodeposition process; during five years of development the approach has enabled void-free, bottom-up, and self-limiting Au deposition with bulk density over a wide range of feature ARs, depths and geometries.[33, 37, 40–44] Here, the limits of this technique are extended to depths approaching 500 µm, obtaining the thickest X-ray masks ever reported in literature, with an AR of 25. The masks were tested in an EI setup with a 160 kVp source, with a dual-threshold photon counting detector used to discriminate the mask's response at different energy thresholds. Validation of the masks was performed by imaging of aluminium (Al) and copper (Cu) wires, obtaining good agreement with simulations based on a wave optics approach. Furthermore, a higher phase sensitivity for the Cu wire was observed when compared to a pair of 300 µm thick masks available from the Nikon X-Tek System.

These results pave the way for higher sensitivity in high-energy XPCI, especially for the denser objects that are often the target of industrial NDT and security inspections, enabling multi-modal imaging.

## 2  Discussion and results

The X-ray masks were designed to have periods of 75 µm and 98 µm with 21 µm and 28 µm wide Si apertures, respectively, for use in an Edge Illumination system (see Methods) accounting for both magnification and the pixel size of the detector. The dimensions allowed for photoresist patterning on the Si substrate by optical lithography, followed by DRIE, with metallization using a sputter-deposited Cr/Au seed layer for electrical conduction, and bottom-up Bi-catalyzed Au filling, shown both schematically and as implemented in **Figure 1**. We used a maskless direct writing laser method (DWL), which allowed us a facile and high-throughput design customization (see Figure S1 in Supplementary Information (S.I.)) to accommodate the different mask dimensions for the trenches and transverse bridges used to support the structure (as well as transverse valleys included to ensure complete wetting). Depending on the complexity of the design and the photoresist (PR) used, the exposure times varied on the order of min/cm². For example, exposures for the 1.25 × 1.1 cm² masks were approximately 7 min. A thick PR was required to maintain masking during the prolonged processes for deep etching; SPR220-7 was spin-coated to a thickness of 6 µm. The thick and robust PR concurrently with relatively wide patterned dimensions allowed for some tolerance in the parameter optimization, with a wide range of exposure intensities and focus heights resulting in satisfactory exposure (see Methods section).

The subsequent etching of the bare Si was carried out using a Bosch process with $C_4F_8$ as the protective deposition precursor and a mixture of $SF_6$ and $O_2$ as the etchants. $C_4F_8$ passivates the substrate by forming a polymer film of $(CF_2)_n$ that protects the underlying Si substrate from etching through the ensuing reaction driven by the fluoride and oxygen radicals.[23] The Bosch process relies on alternating flows of the protective and etching gases, controlling their amount, flow time, and plasma power during their flow. Adjustments in the parameters for $SF_6$ or $C_4F_8$ promote either etching or isotropic deposition of a protective layer. Preferential vertical etching is achieved by applying a RF vertical bias, controlled by the 'platen power' during the etching steps, which transfers kinetic energy to the etching gases to promote physical sputtering at the bottoms of the trenches.

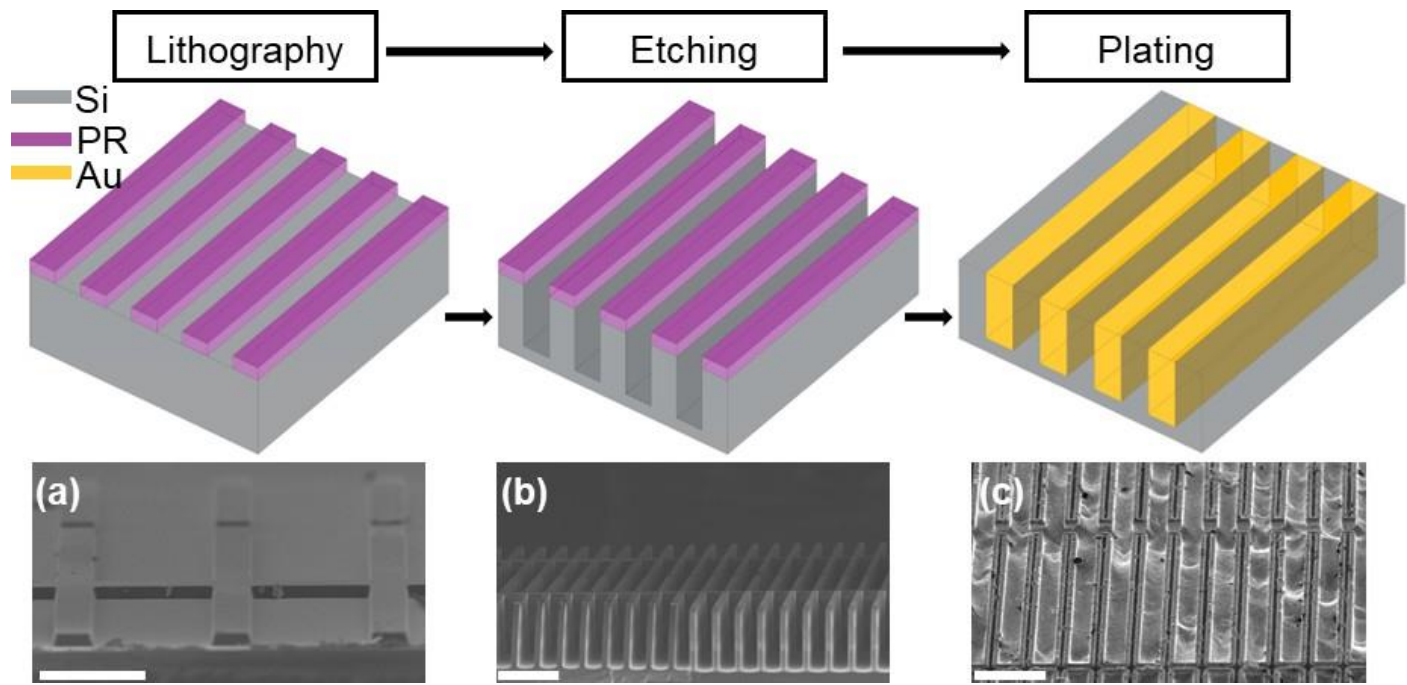

Figure 1: Schematic and implementation of the fabrication process. a) SEM bird-eye view of a fully developed SPR220-7 mask, b) SEM bird-eye view of a partly etched mask, c) SEM bird-eye view of a fully Au plated (detector) mask. Scale bar: a) 50 µm, b-c) 100 µm.

Table 1: DRIE parameters during optimization

|    | $C_4F_8$ time (s) | $SF_6$ time (s) | Platen (W) | $C_4F_8$ amount (sccm) | $SF_6$ amount (sccm) | N cycles |
|----|---|---|---|---|---|---|
| S1 | 6 | 8 | 12 | 85 | 130 | 470 |
| S2 | 7 | 9 | 20 | 100 | 150 | 500 |
| S3 | 7 | 9 | 20 | 120 | 150 | 200 |
| S4 | 8 | 9 | 20 | 100 | 150 | 200 |
| S5 | 7 | 9 | 16 | 100-85 | 150-130 | 200 |

The photoresist SPR220-7 exhibited a selectivity of ≈ 100:1 compared to etching of bare Si, with the 6 µm coating thus allowing selective etching to depths beyond the 500 µm target value. The first set of parameters (S1 in Table 1) resulted in a loss of anisotropy after 250 cycles, corresponding to a depth of 200 µm (see **Figure 2a**). Hence, a second set of parameters (S2) was chosen to promote reactant transport diffusivity and vertical momentum by increasing the cycle duration, gas amount, and platen power. These parameters created a harsher environment with a higher etching rate (≈ 1 µm/cycle), resulting in depths as high as 600 µm but at the cost of extreme thinning of the lamellae (see **Figure 2b**). To mitigate this effect, samples were etched under the same conditions but with augmented exposure to $C_4F_8$ by either 1 s or 20 sccm to increase the thickness of the protective layer to protect the sidewalls of the Si lamellae. This test showed that adding 20 sccm (S3, see **Figure 2c**) maintained visible bowing of the lamellae even at 200 µm depth, while adding 1 s to the exposure time (S4, see **Figure 2d**) visibly reduced this bowing. Micro-masking, as visible in the black Si forming in Figure 2c-d, due to masking by adventitious contaminants or residuals of passivation layer can be reduced by descum in $O_2$ plasma (see Methods). Black Si formed especially when the samples were not properly adhering to the carrier wafer, likely due to uneven exposure to processing conditions that enhanced the micro-structuring effect. Reducing platen power to 16 W for 200 cycles resulted in slightly negative tapering (S5, see **Figure 2e**), even with reduced gas flow (not shown), the lower $C_4F_8/SF_6$ time ratio promoting sidewall etching over protection. The results from S1 to S5 illustrate the significant impacts of platen power and cycle durations on the etching process, with gas flow having a lesser impact and therefore not considered in the final process optimization.

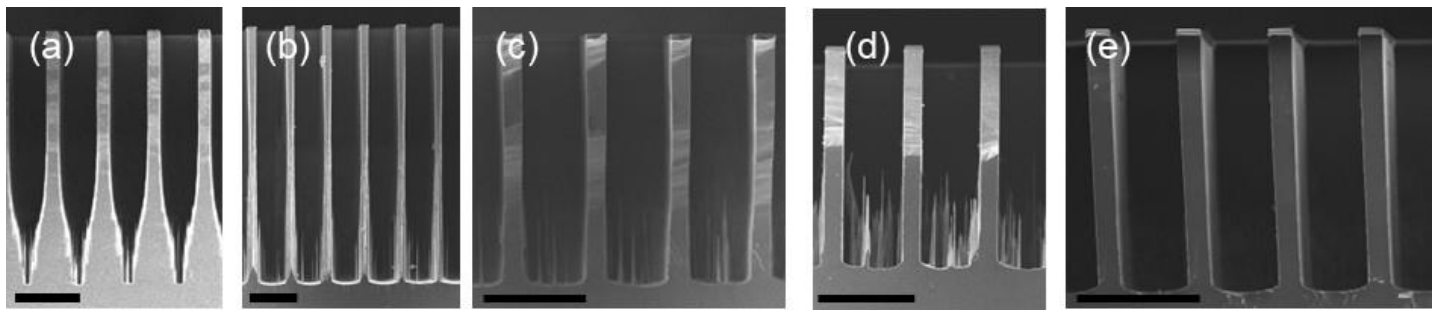

Figure 2: a) Etching result with initial parameters (S1), b) harsher parameters (S2), c) 20sccm C4F8 (S3), increased time (S4) and reduced platen power (S5). Scale bar: 100 µm

Ultimately, by proportionally increasing the $C_4F_8$ and $SF_6$ exposure times, diffusion towards greater depths was promoted, pushing the etching downward. The platen power was also increased to provide more kinetic energy to the gaseous reactants, thereby increasing the etching rate at greater depths, albeit at the cost of increased bowing due to physical sputtering. Balancing these two parameters as a function of trench depth was essential for achieving greater depths while minimizing bowing. A stepped approach was adopted, adjusting the parameters progressively, increasing cycle duration and platen power, and balancing them with the appropriate number of cycles, as shown in **Figure 3**. The optimized parameters for approximately 1 cm² samples are detailed in Table 2: platen power was increased from 12 W to 20 W in 2 W steps every 130 cycles, and the durations of both deposition and etching steps in each cycle were extended together to slightly increase the $C_4F_8/SF_6$ ratio and reduce bowing. Figure **3a** and **Figure 3b** show the intermediate profiles of the lamellae at approximately 150 µm and 300 µm, respectively. Widening of the lamellae toward the bottom was preserved to balance the more severe lateral etching required for Steps 4 and 5. **Figure 3c** shows that bowing is reduced and there is only modest widening at the bottom of the lamellae at the end of processing, acceptable for our applications and possible to further fine tune if necessary.

Table 2: Optimised parameters for ramped parameters etching

|  | $C_4F_8$ time (s) | $SF_6$ time (s) | Platen (W) | $C_4F_8$ amount (sccm) | $SF_6$ amount (sccm) | N cycles |
|---|---|---|---|---|---|---|
| Step 1 | 6 | 8 | 12 | 130 | 85 | 130 |
| Step 2 | 6 | 8 | 14 | 130 | 85 | 130 |
| Step 3 | 7 | 9 | 16 | 130 | 85 | 130 |
| Step 4 | 7 | 9 | 18 | 130 | 85 | 130 |
| Step 5 | 7.5 | 9.5 | 20 | 130 | 85 | 100 |

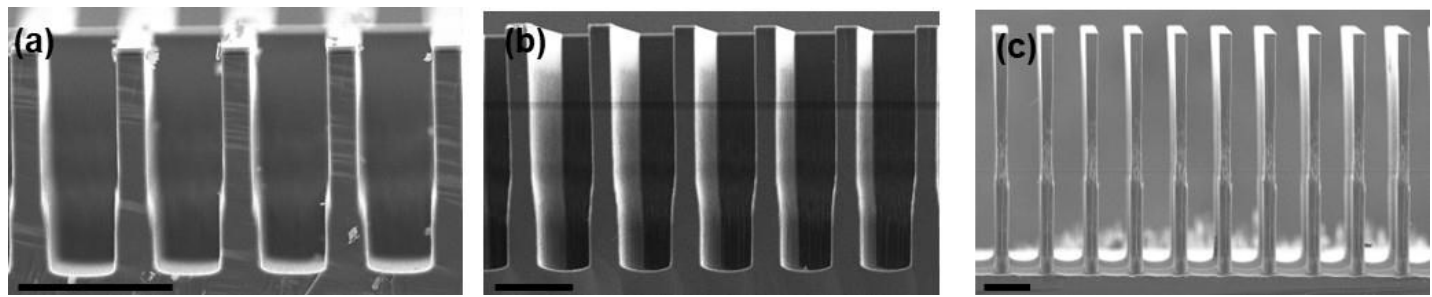

Figure 3: SEM front view images of the masks etched by stepped parameters etching showed in Table 2. a) results after Step 2, b) after Step 3, c) final results after Step 5. Scale bar: 100 µm

An alternative route via wet MACE was also attempted. The selection of liquid solution rather than vapor-phase MACE was dictated by the more complex experimental set-up required for the latter. Here, a 2.5 µm thick bi-layer PR coating constituted of LOR (≈ 800 nm) and S1818 (≈ 1.7 µm) provided significant undercut that allowed the subsequently sputter-deposited catalyst Ti/Au layer to completely adhere to the Si substrate (see Methods). The characteristics of the adhesive and catalyst layers have a significant impact on the etching process. In particular, e-beam evaporated Ti/Au was effective in enabling MACE, whereas the thermally evaporated Cr/Au layer failed to initiate the etching. Additionally, a 2-minute oxygen plasma treatment prior to evaporation appeared to be crucial for the MACE process to proceed effectively, likely due to the resulting higher wettability of the underlying Si. The etching reaction was carried out in an aqueous solution of $H_2O_2$ and HF, starting with a solution comprised of HF, $H_2O_2$, and $H_2O$ with a 5.3/0.25/50 molar ratio, as in Ref. 30. There, the Chartier factor and the temperature were kept rather low to accommodate the slower rate appropriate for vertically etching nanoscale features. Here, because of the catalyst's two orders of magnitude larger lateral size, the metal layer fractured regardless of the Chartier factor, the introduction of EtOH in solution as a surfactant, and the temperature used (see Methods and Figure S5 in S.I.). To improve the metal layer's stability, reduction of the longitudinal dimension and introduction of transverse bridges to reduce mechanical stress, as well as increase of Au thickness to 20 nm from 10 nm were tried. The former did not significantly improve the metal's stability, and the latter completely inhibited etching due to reduced transport through the thicker layer.[45, 46] In these large and wide features, catalyst stability seems challenging to maintain even at low ARs due to unbalanced Van der Waals forces between Si and Au.[28] DRIE, in contrast, as already shown, required only modest process optimization to achieve sufficiently vertical etching.

Au electrodeposition in the dry-etched Si trenches requires a wettable, electrically conductive ("seed") layer due to the high resistivity of Si as well as poor nucleation and wetting in the electrolyte. A conformal seed layer on all surfaces, including the field over the trenches as well as the trench sidewalls and bottom, can be obtained by chemical deposition such as ALD which guarantees a self-limiting layer-by-layer deposition.[32, 34] However, magnetron sputtering is a quicker and less costly alternative for a continuous layer even at AR > 20, due to its relatively high operating pressure as compared to other PVD techniques, which results in a more isotropic deposition and an improved step coverage.[30] For relatively small patterned areas, the mask can also be optimized to improve transport of the gaseous reactant inside the trenches by etching the region around it, to create protrusive lamellae. The external region, or 'frame,' also facilitates specimen handling and, for sufficiently deep features, may provide shorter paths for electrical conduction to the bottoms of the Au-sputtered trenches. A 2 mm wide frame was utilized in this work (see Figure 1 in S.I.).

Both a Cr adhesion layer and the Au seed layer were sputtered at relatively low pressures, 5 mTorr (0.67 Pa) and 3.75 mTorr respectively, to obtain higher deposition rates, at the cost of some decrease of step coverage.[47–49] The trenches were coated with approximately 150 nm Cr and 150 nm Au; the coating was thinner inside the features but still continuous. Conventional electrochemical deposition with a static working electrode and commercial sulfite-based Au electrolyte was attempted. Despite vigorous stirring and bubbling throughout the whole duration of plating to force convection inside the trenches, and a wide range of plating conditions used (see Methods), Au was deposited preferentially at the top of the lamellae, eventually clogging the apertures and blocking the plating inside the trenches.

The development of a $Bi^{3+}$-catalyzed bottom-up Au filling electrodeposition process and its extension to high aspect ratio trenches and vias in gratings for advanced X-ray imaging technologies, including gratings patterned across 100 mm wafers, has been presented in a series of publications.[33, 35, 37, 40–44, 50–52] Simulations, based on a mechanistic model, capture key experimental observables of the process in the $Bi^{3+}$-containing sulfite electrolytes.[35] The electrolytes and processes provide dense, void-free Au fill in Au/Si gratings, including deep and high aspect ratio features needed in X-ray interferometry and imaging applications. Gratings with trenches of aspect ratio (depth/width) exceeding 60 and gratings with trenches as deep as 305 μm, have been filled void-free across 100 mm Si wafers, as have intentionally curved gratings.[40, 41, 43, 51, 52] Function of the gratings (assessed by visibility as well as X-ray phase contrast imaging (XPCI) of biological samples) is also consistent with fully dense and void-free Au fill. The bottom-up Au filling process was successfully extended to the 500 μm deep trenches of this study.

Gold electrodeposition was conducted at room temperature in a three-electrode electrochemical cell containing 400 mL of electrolyte composed of 0.16 mol/L $Na_3Au(SO_3)_2$ + 0.64 mol/L $Na_2SO_3$ and containing

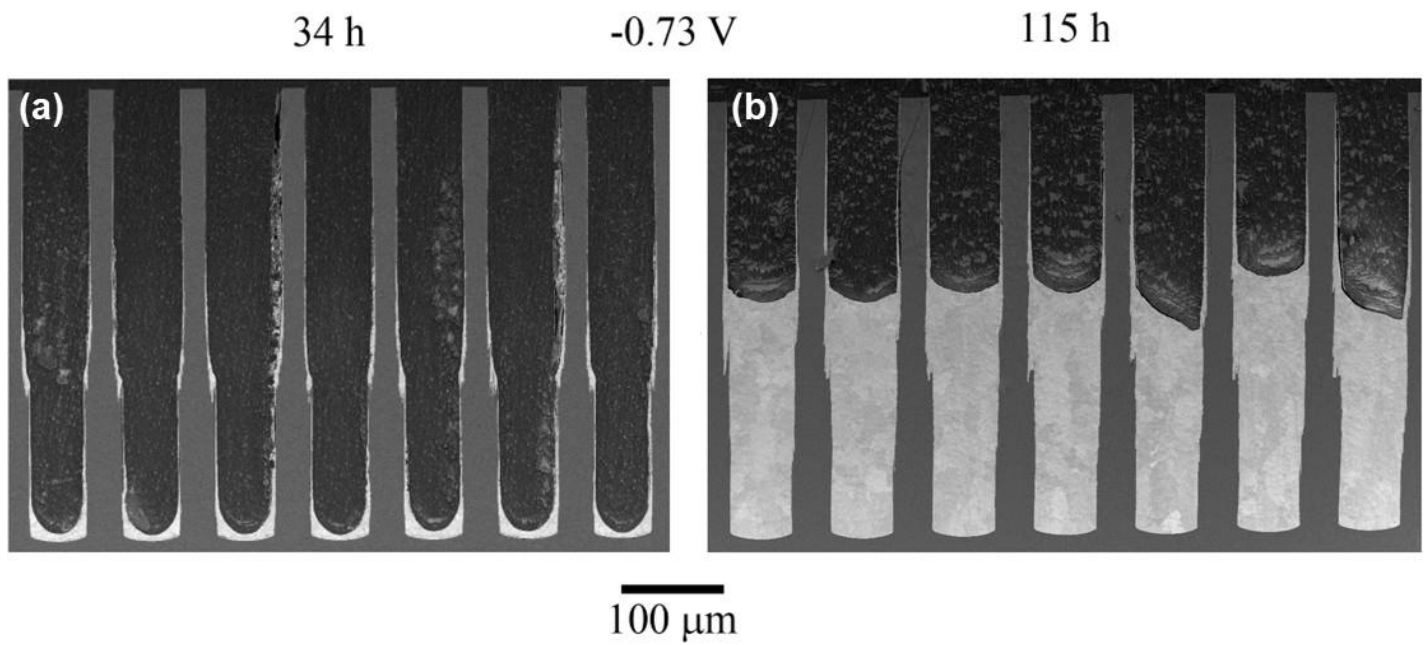

Figure 4: Cross-sectioned gratings with 98 µm pitch 500 µm deep trenches after Au deposition, imaged by optical microscope; higher (4×) magnification images shown in the top row capture the bottom-up filling evolution while the lower image captures the modest variation of the half-filled grating at a longer length scale. Depositions times and applied potentials relative to SSE are as indicated. The specimens were filled at room temperature and rotation rate of 200 rpm.

50 µmol/L $Bi^{3+}$. Electrolyte pH was maintained at the near neutral value of 9.0. Potentials were measured relative to a $Hg/Hg_2SO_4$/saturated $K_2SO_4$ reference electrode (SSE) separated from the main cell by a Vycor fritted bridge. The electrolytes were derived from 0.32 mol/L $Na_3Au(SO_3)_2$ source solution (Technic Gold 25-F replenisher concentrate) and $Na_2SO_3$ salt in 18 MΩ·cm water. The $Bi^{3+}$ additive was introduced by anodic dissolution of 99.999 mass % fused Bi metal electrode on the assumption of a 100 % efficient Bi → $Bi^{3+}$ + 3e⁻ dissolution reaction at -0.45 V SSE making the stated $Bi^{3+}$ concentration an upper bound. Sections of a grating with 500 µm deep trenches that were partially filled with Au at the indicated potential are seen in cross-section in **Figure 4**. Modest variation of the Au-fill, such as is seen across the fully filled grating in Figure 1c, is evident in the cross-section in **Figure 4b**. The potential applied during deposition in the gratings used for imaging ranged from -0.73 V to -0.78 V in order to both accelerate deposition and achieve more complete filling of the trenches.

The masks were tested in an EI setup characterized by a conventional X-ray source with a tube voltage up to 160 kVp and a photon-counting detector. The latter featured two counting thresholds. The first was used to cut off noise, while the second enabled splitting of the detected spectrum into two parts, thereby allowing some degree of discrimination of the incoming photons' energies, following the scheme in **Figure 5**. In EI, one of the masks is placed just upstream of the sample ('sample mask') and the second mask is placed very close to the detector ('detector mask'). The periods and apertures of the masks were chosen as appropriate for the system magnification (see Methods) and the 100 µm pixel size of the detector such that, projected on to the detector plane, each period would cover exactly one pixel. The intensity transmitted to the detector varies with the relative lateral displacements of the masks, exhibiting a gaussian-like behaviour, referred to as an Illumination Curve (IC). The IC is used to retrieve the attenuation, phase shift, and scattering induced by the sample. The IC peak is obtained when the lamellae of the masks are aligned so that the maximum intensity is transmitted to the detector (**Figure 5a**), whereas the baseline is obtained when they are completely misaligned so that the beamlets created by the sample mask are absorbed by the Au-filled trenches in the detector mask (**Figure 5c**). When a sample is placed between the two masks, the IC's area is reduced due to the sample's attenuation, its center is laterally shifted due to refraction, and its Full Width at Half Maximum (FWHM) is increased by the sample's small angle scattering. The point on the IC with the highest sensitivity to refraction, i.e., to lateral shift of the IC, is that corresponding to the steepest slope (**Figure 5b**). This is obtained when the two masks are displaced by roughly half the size of the aperture of the masks, so that the transmitted intensity is midway between the IC peak and baseline values. This relative positioning of the masks was utilized herein to assess the maximum phase sensitivity achievable by the system. The images were taken at 160 kVp, and the response at different energies was determined by setting the detector's energy thresholds at 25 keV, 45 keV, 65 keV, or 85 keV. Increasing the energy threshold results in fewer photons being counted, reducing the signal-to-noise ratio (SNR) in the images. This effect was worsened by the lower flux emitted by the source at higher energies; longer exposure times were therefore used to improve the statistics. Images of 1.5 mm diameter Al and 1.0 mm diameter Cu wires were obtained with 12 µm relative displacements of the masks to meet the highest sensitivity requirement. To increase the image resolution, the wires were scanned in front of the sample mask in steps of either 1.5 µm or 3.8 µm, depending on the flux available. To reduce the overall number of required scans while still obtaining the desired number of energy windows, both the high and low energy thresholds were simultaneously exploited, namely in a scan with lower and higher thresholds at 25 keV and 65 keV, respectively, and another scan using 45 keV and 85 keV. Because the 45 keV to 85 keV scan resulted in a lower flux than the 25 keV to 65 keV scan, a smaller step of 1.5 µm was employed (as opposed to the 3.8 µm step used for the 25 keV to 65 keV scan) to achieve overall comparable statistics (per binned pixel).

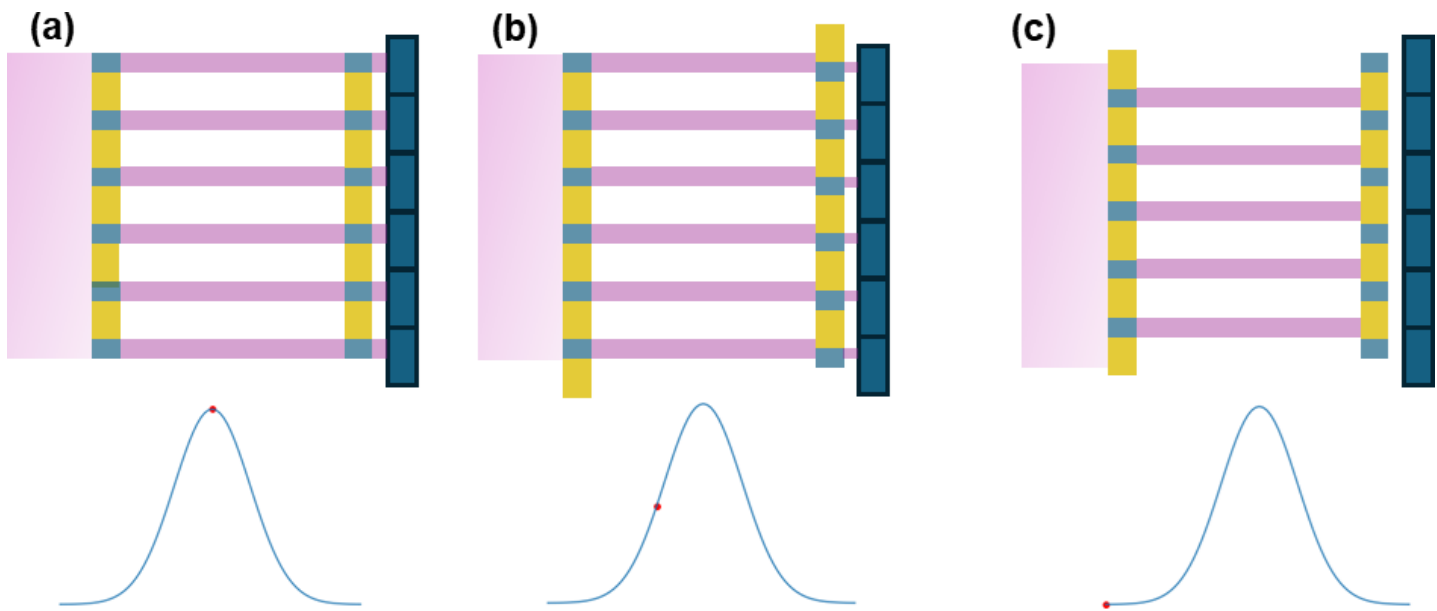

Figure 5: Top row: Edge Illumination scheme at different masks positions. The X-ray beam is purple, the masks' opaque gold septa yellow, the Si lamellae are light blue. In dark blue, the detector pixels. Bottom row: The corresponding value on the Illumination Curve is highlighted by the red dot.

Due to their simple geometry, wires are commonly used to quantitatively test the response of XPCI systems. The phase sensitivity is assessed using the refraction-induced edge enhancement, manifest as a characteristic brighter (or darker) edge. The described set-up was also simulated using a wave optics approach simulating a set of masks with 500 µm thick Si lamellae and 470 µm thick Au septa.[53] The simulated refracted intensity at the edge of the Al wire is compared to the experimental results for every energy threshold examined. As shown in **Figure 6a**, theory and experiment are in good agreement, further evidence that a gold deposit with bulk-like density was obtained following superconformal Au filling and that all the fabrication requirements were met. The Cu wire's refraction intensity was, however, found to be lower experimentally than in the simulation (not shown). This may be caused by parasitic transmission through the Au septa of higher energy photons. Specifically, the hardened fraction of the beam that is transmitted is not attenuated by lower density samples like the Al wire, but the more highly absorbing Cu wire is able to significantly attenuate even these higher energy photons transmitted through the Au septa.

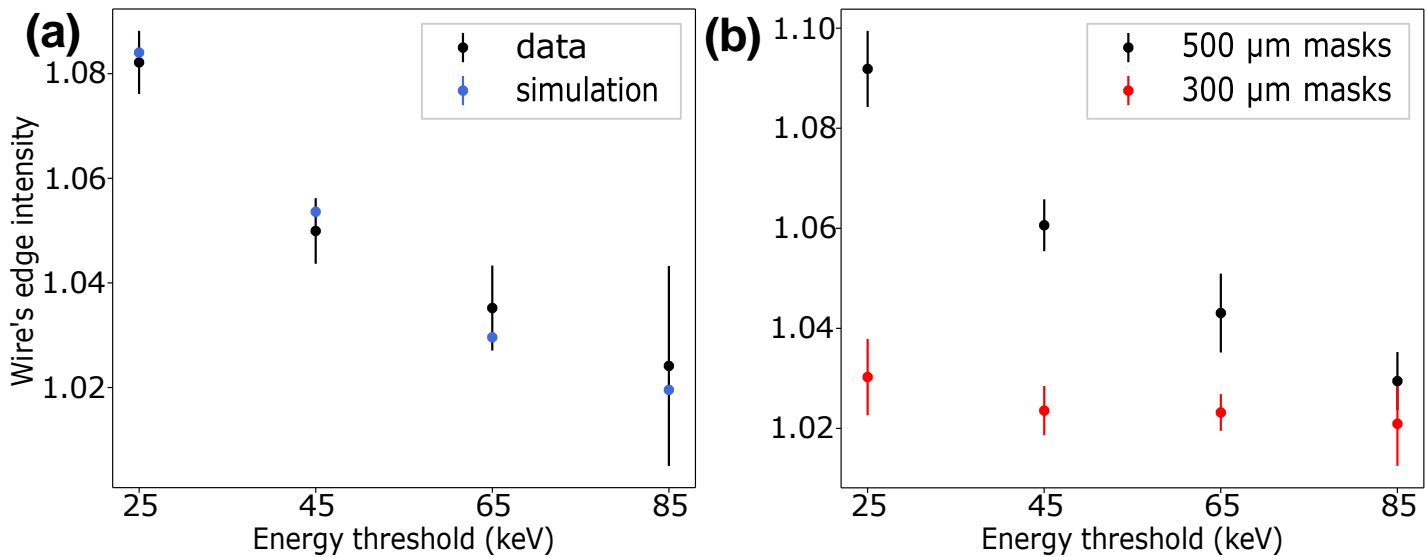

Figure 6: a) Al wire's edge enhancement obtained with 500 μm masks (black) and by wave optics simulation (blue). b) Cu wire's edge enhancement obtained with 500 μm masks (black) and with 300 μm masks (red). Error bars represent one-standard deviation.

The consequence is a blurring of the image and a reduction of the refraction intensity, which worsens when thinner masks are used. In fact, when a set of 300 μm masks is used, the refraction intensity is significantly reduced almost irrespective of the threshold chosen, as shown in **Figure 6b**. The superior performance achieved by the thicker masks, which lead to up to 3 times the phase sensitivity compared to the 300 μm masks, shows their potential for high-energy XPCI of dense/high-Z materials. It is noteworthy that performance can be improved by eliminating the transverse Si and Au bridges which currently limit the vertical field of view available with these masks. While this problem could be overcome by scanning the sample, this would also lead to longer acquisition times, making it more convenient to enlarge the field of view by increasing the masks' size. Critically, both etching and plating showed good homogeneity across the mask, which holds promise for the fabrication of larger structures.

## 3    Conclusions

A multi-step fabrication process including maskless photolithography, dry etching, and Au bottom-up plating for ≈ 500 μm thick Si and Au X-ray masks has been optimized for an EI imaging setup. DRIE showed a significantly higher compatibility to the etching of the tens-of-micrometers wide structures required in this case compared to MACE, whose stability was impacted by fracture of the catalyst during the process. The lamellae's verticality was improved by dynamically changing the dry etching parameters in a Bosch process to compensate for the increasing AR. To maximize the masks' attenuation, a highly dense Au deposit was obtained by Bi-catalyzed Au electrodeposition. This additive-derived bottom-up filling process was successfully used for the first time on gratings of such depth, which are also the thickest X-ray masks used for imaging to date. This result enabled XPCI at 160 kVp with higher phase sensitivity compared to previous x-ray masks, while using a readily available commercial X-ray system. The fabrication of ultra-thick masks could pave the way to high-energy XPCI helping to close the gap between phase-based X-ray imaging and those areas of non-destructive investigations where dense and high-Z materials need to be examined, e.g., industrial inspection, damage assessment, cultural heritage scans, among other applications.

# 4 Methods

*Photoresist coating* :
The substrate used was a 100 mm (4") Si wafer (p-type, boron-doped, <100>, (1 to 10) Ω cm, 800 μm thick; MicroChemicals). It was cleaned for 2 min in acetone and isopropyl alcohol and for 10 min in oxygen plasma, followed by a 2 min dehydration anneal at 150 °C. A layer of HMDS was spin-coated onto the wafer to promote adhesion with the photoresist. A bi-layer photoresist was used to obtain an undercut for MACE. An underlayer (LOR10B, Kayaku Advanced Materials) and the photoresist S1818 (MICROPOSIT Series, Kayaku Advanced Materials) were chosen, spun at 4000 rpm and soft baked at 115 °C for 1 min. The photoresist used for DRIE, SPR 220-7 (MEGAPOSIT Series, Kayaku Advanced Materials), was spin-coated at 4000 rpm (8000$\pi$ rad/min) for 40 s after an initial spin at 500 rpm for 2 s for more homogeneous wetting. Both values were reached by a controlled 2 s long ramp. It was subsequently soft baked for 2 min at 90 °C and for 3 min at 115 °C

*Optical Lithography (design)*:
The photoresist patterning was obtained by Direct Write Laser (DWL 66+, Heidelberg Instruments). For SPR 220-7, the DWL exposure was tuned to a power of 70 mW set at 80 % intensity. The resulting photons were filtered by 50 % and the process was repeated three times to obtain a satisfactory exposure. We note that changing the filter and intensity by ±20 % and the focus offset by ±5 % led to negligible differences. A post-exposure bake equal to the soft bake was done on the exposed wafer. Once cooled, the wafer was developed in MF-319 (Kayaku Advanced Materials) for 270 s leaving no residual layer. The designs to be exposed were prepared using the software KLayout.

*Metal Assisted Chemical Etching*:
The metal catalyst was electron beam evaporated on the bi-layer PR pattern at a deposition rate of 0.1 nm/min. Different thicknesses of Ti and Au were obtained by adjusting the duration of deposition. The PR lift-off was performed in dimethyl sulfoxide (DMSO) in a sonicating bath at 65 °C. MACE was conducted following the procedure in ref. 30 and the $H_2O_2$/HF ratio was increased by 2, and 5 times either at room temperature or keeping the bath at ≈ 7 °C. Both HF and $H_2O_2$ were dissolved in water at, respectively, 48 % and 30 % w/w concentration. Therefore, only the excess water needed to attain the desired concentrations was added to the etching bath. This excess water was also replaced by pure EtOH as a surfactant, systematically varying the $H_2O_2$/HF ratios by 2, 4, 6, 8, 12, 16, and 24 times.

*Deep Reactive Ion Etching*:
Prior to any dry etching, the wafer was cleaned in $O_2$ plasma for 10 min to dissolve adventitious contaminations resulting in micro-loading, as partially visible in Figure 2d. The instrument used was an STS DRIE (STS ICP ASE) System, configured for Bosch processes. Longer processes (time > 1 h) were intermittently halted to prevent sample overheating. Also, the chamber was conditioned for 25 min to the Step 1 conditions in Table 2 before every deposition to ensure a controlled initial environment.

*Conventional plating* :
To prepare the Si gratings for Au electrodeposition, a conductive layer was sputtered by PVD (Lesker PVD75 Sputter Coater system). The Cr adhesive layer was sputtered for 15 min in 5 mTorr (500 W, 11 nm/min) and Au for 20 min in 3.75 mTorr (50 W, 7 nm/min). Conventional plating in NB Semiplate Au 100 (Arsenic-based, MicroChemicals) bath was attempted using currents of 1 mA/cm² to 6 mA/cm², direct current as well as pulsed, in conditions of low and vigorous stirring, using a stationary electrode. The results showed consistent clogging of the aperture, although vigorous stirring promoted plating to lower depths.

*X-ray Nikon systems*:
The X-ray system was arranged in a stand-alone lead-shielded cabinet at the Nikon Metrology UK's factory in Tring, Hertfordshire, UK, including a Tungsten X-Tek 160 tube with a focal spot of approximately 100 μm as the X-ray source. The detector was a dual-energy single photon counting CdTe CMOS (XCounter XC-FLITE FX2) with 2048 × 128 square pixels with 100 μm sides, placed 2 m from the X-ray source.

The two masks ("object" and "detector") were placed at 1.5 m and 1.95 m from the source, so that their projected periods would match the detector pixels (i.e., in the detector plane, due to geometrical magnification, their actual pitches of 75 μm and 98 μm were magnified to be 100 μm). The two masks were aligned following a moiré fringes approach,[54] and prior to every scan, the object mask was laterally shifted by small fractions of the grating pitch across the IC to find its FWHM. The wires were scanned just downstream of the object mask with the lateral resolution defined by the scan speed and the pixel binning. The detector acquisition rate was set to 1 fps or 2 fps (frames per second) and the wires were scanned over one object mask's period in 20 and 50 dithering steps (respectively 3.8 μm and 1.5 μm long) for, respectively, the 25 keV to 65 keV and the 45 keV to 85 keV scans. The experimental values shown in Figure 6 are found by averaging the edge enhancement over the pixel rows with higher visibilities for both sets of masks to eliminate results aligned with, and thus impacted by, the transverse Si and Au bridges.

*EI experiments simulations and validation*:
An EI system was simulated based on the work published by Vittoria et al.,[53] where an angular spectrum approach is used to propagate the polychromatic X-ray beam through the masks and wires in the system. The spectrum was calculated by the Spekpy 2.0.8 online toolkit inputting the source characteristics in the real set-up (E = 160 kVp, current = 4.4 mA/s anode angle at 11° and (0.8 ± 0.1) mm Be window). The detector's response was included by calculating the CdTe sensor's absorption using data from Xraylib library and its energy thresholding was obtained by completely filtering out the energies off the entire spectrum. The masks' and wires' complex transmissions were included at the appropriate distances from the source (whose width was set to 100 μm size) and the detector using values found in Xraylib.[55] The masks' and wires' geometrical sizes were simulated in the real space as an ideal square wave and cylinder.

## Supporting Information
Supporting Information is available from the Wiley Online Library or from the author.

## Acknowledgements
This work was funded by EPSRC (grant EP/T005408/1). The work was also supported by the European Research Council, ERC-StG-IntelGlazing grant no. 679891. A.O. was supported by the Royal Academy of Engineering under the "Chairs in Emerging Technologies" scheme (grant CiET1819/2/78). Certain equipment, instruments, software, or materials are identified in this paper in order to specify the experimental procedure adequately. Such identification is not intended to imply recommendation or endorsement of any product or service by NIST, nor is it intended to imply that the materials or equipment identified are necessarily the best available for the purpose. The authors also acknowledge the assistance of the technical team in the London Centre for Nanotechnology (LCN): Steve Etienne, Vijayalakshmi Krishnan, Lorella Rossi, Rohit Khanna, and Suguo Huo.

**Table of Contents**

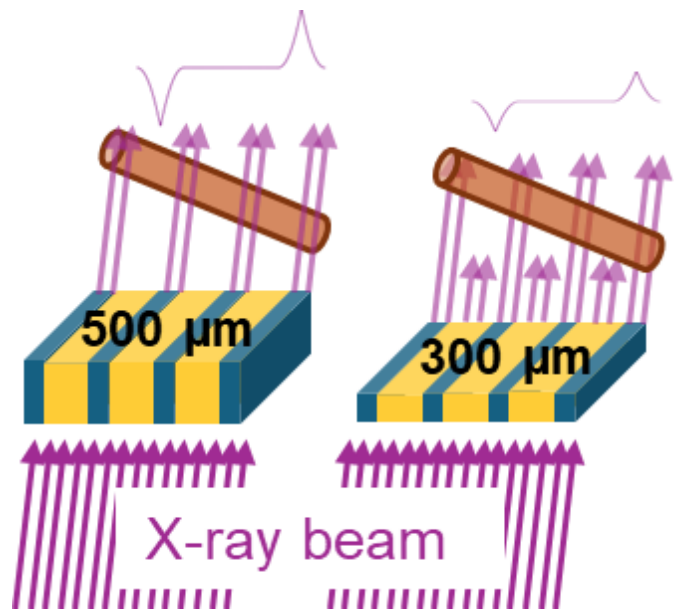

This work shows a fabrication route to 500 µm thick X-ray masks - the thickest ever reported in literature - combining dynamic optimization of etching parameters and superconformal bismuth-mediated gold filling. Their use in an Edge Illumination set-up showed their higher phase sensitivity at higher energies compared to thinner masks, especially for denser materials like copper.